\documentclass[journal]{IEEEtran}
\usepackage{blindtext}
\makeatletter
\newcommand\xleftrightarrow[2][]{%
  \ext@arrow 9999{\longleftrightarrowfill@}{#1}{#2}}
\newcommand\longleftrightarrowfill@{%
  \arrowfill@\leftarrow\relbar\rightarrow}
\makeatother

\usepackage{amsmath}
\usepackage{graphicx}
\usepackage{multirow}
\usepackage{cite}
\usepackage{array}
\usepackage{kantlipsum}
\usepackage{url}
\usepackage{mathtools}
\usepackage{eqnarray}

\usepackage{xcolor}
\newcolumntype{C}[1]{>{\centering\let\newline\\\arraybackslash\hspace{0pt}}m{#1}}
\usepackage{wrapfig}
\usepackage{makecell}
\usepackage{multirow}
\usepackage{multicol, blindtext}
\usepackage{xcolor}
\usepackage{tikzpagenodes}

\usepackage[a4paper,margin=1in]{geometry}
\usepackage{fancyhdr}
\begin{document}

\title{Secure Distribution of Protected Content in Information-Centric Networking}

\author{Muhammad~Bilal,~\IEEEmembership{Member,~IEEE,}
Sangheon~Pack,~\IEEEmembership{Senior~Member,~IEEE}
\thanks{This work is supported in part by the Hankuk University of Foreign Studies research funds for 2019 and in part by National Research Foundation (NRF) of Korea Grant funded by the Korean Government (MSIP) (No. 2017R1E1A1A01073742).}   
\thanks{Muhamamd Bilal is with Hankuk University of Foreign Studies, Yongin-si, Rep. of Korea. (email: m.bilal@ieee.org)}
\thanks{Sangheon Pack is with Korea University, Seoul, Rep. of Korea. (email: shpack@korea.ac.kr)}
\thanks{A preliminary version of this paper was presented at IEEE VTC 2018-Spring RAFNET Workshop, Porto, Portugal, June 2018~\cite{Bilal18VTC}.}}

\maketitle
  \begin{tikzpicture}[remember picture,overlay]
    \node[align=center,text=red] at ([yshift=1em]current page text area.north) {This article is an  enhancement version of journal article published in IEEE Systems Journal, DOI: 10.1109/JSYST.2019.2931813};
    \node[align=center,text=blue] at ([yshift=-1em]current page text area.south) {1937-9234 © 2019 IEEE. Personal use is permitted, but republication/redistribution requires IEEE permission.};
  \end{tikzpicture}%

\begin{abstract}
 The benefits of the ubiquitous caching in ICN are profound, such features make ICN promising for content distribution, but it also introduces a challenge to content protection against the unauthorized access. The protection of a content against unauthorized access requires consumer authentication and involves the conventional end-to-end encryption. However, in information-centric networking (ICN), such end-to-end encryption makes the content caching ineffective since encrypted contents stored in a cache are useless for any consumers except those who know the encryption key. For effective caching of encrypted contents in ICN, we propose a secure distribution of protected content (SDPC) scheme, which ensures that only authenticated consumers can access the content. SDPC is lightweight and allows consumers to verify the originality of the published content by using a symmetric key encryption. Moreover, SDPC naming scheme provides protection against privacy leakage. The security of SDPC was proved with the BAN logic and Scyther tool verification, and simulation results show that SDPC can reduce the content download delay. 
\end{abstract}

\begin{IEEEkeywords}
Information Centric Networking, Named Data Networking, 5G, Content Distribution, Access control, In-network caching, Security, Authentication, Privacy.
\end{IEEEkeywords}

\IEEEpeerreviewmaketitle

\section{Introduction}
\label{sec:intro}
Since the earliest time of the Internet, its underlying architecture has been based on packet-switching and host-to-host communications. The TCP/IP layered architecture employs the same view and provides an abstract host-to-host communication model to communication applications. It decouples what to communicate from how the communication is done. This basic design feature of the TCP/IP architecture was far-reaching, allowing the Internet to grow for almost four decades while adopting various features and yet maintaining high efficiency. However, in the recent past there has been a profound increase in Internet connectivity, and with the emergence of new Internet applications, the Internet semantics have changed from host centric to content centric. To satisfy the needs of emerging internet applications, the current TCP/IP Internet architecture has adopted several application layer solutions known as over-the-top (OTT) applications, such as content delivery network (CDN), web caching, and peer-to-peer networking~\cite{Ge13, Malatras15}. With the addition of numerous applications, the gap between the basic semantics of the current Internet architecture and its usage is bound to increase; in fact, the additions of new OTT applications are leading us towards a very complex Internet architecture, and are introducing challenges to achieving efficiency, security, and privacy at acceptable economical cost.

Further, Internet trends are shifting away from browsing information to online consuming and sharing all types of content, including user-generated contents. Hence, the most promising characteristic of the future Internet is ubiquitous content delivery. What is being communicated is becoming more important than who is communicating. In this perspective, information-centric networking (ICN) has emerged as a promising architecture for the Future Internet; recently, the ICN support for 5G use cases were specified by NGMN. ICN represents a paradigm shift from host-centric to content-centric services and from source-driven to receiver-driven approaches. In the ICN paradigm the network is in charge of doing the mapping between the requested content and where it can be found. To do so, a network level naming is used for identifying content objects, independent of their locations~\cite{Jacobson09,Ahlgren12,Zhang13}. This means that the ICN architecture decouples contents from the host at the network level and supports a temporary storage of contents at in-network caches.

In ICN, in-network caching is an integral part of the ICN service framework~\cite{Jin12}\cite{Bilal17Access}. The benefits of the ubiquitous caching in ICN are profound, but it also introduces a challenge to content security; especially, the protection of a private or confidential content is a challenging task.  The ICN enabled cache routers can store the content segments for future use; hence, the content is temporarily cached in few intermediate cache routers while it is being delivered to a consumer. If content requests traverse a cache router that holds a temporarily cached copy of that particular content segment, then the request is entertained locally without being routed towards the publisher. However, in ICN the publisher has no control over the content after injecting it in the network; in particular, if  a private or confidential content is protected insecurely, then any unauthorized consumer can acquire it from intermediate caches. Traditionally, the protection of the content against unauthorized access requires consumer authentication and involves the conventional end-to-end encryption. Consequently, when the content is encrypted with the authorized user's key, the in-network caching becomes ineffective in ICN. 

\begin{figure}
\centering
\includegraphics[width=0.42\textwidth]{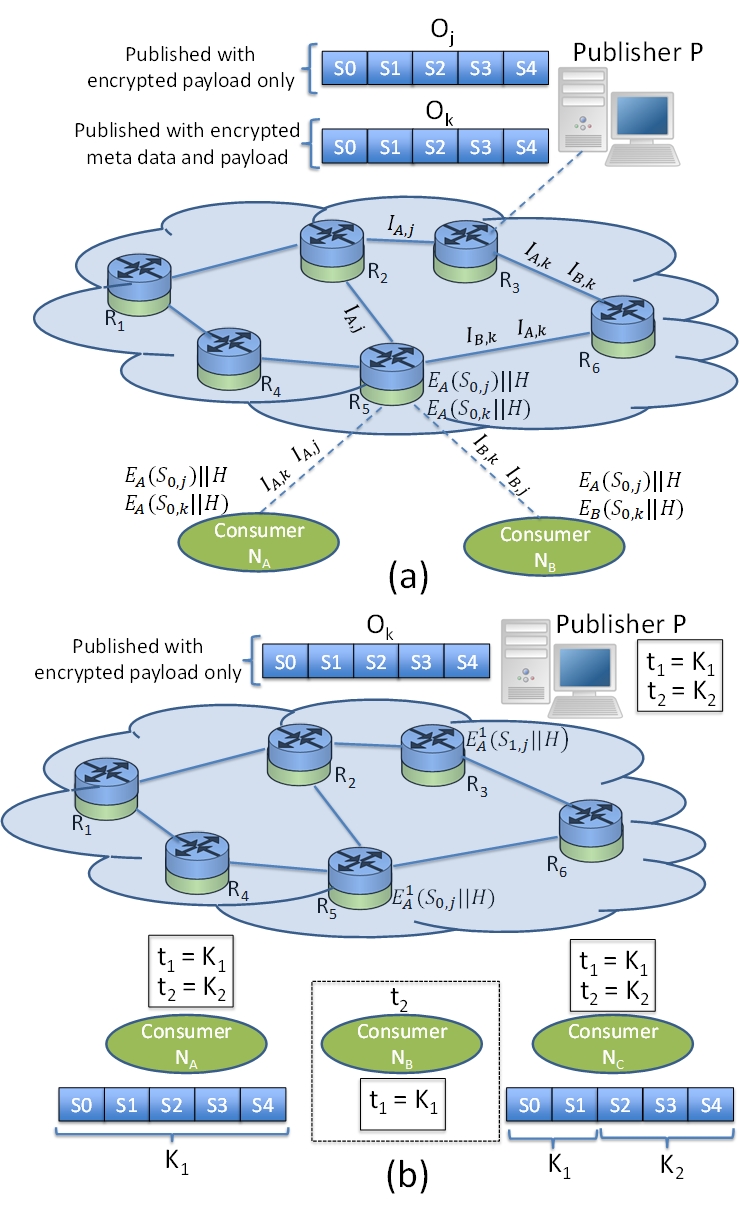}
\caption{Ineffective caching in ICN with end-to-end encryption.}
\label{fig:1}       
\end{figure}

In Figure~\ref{fig:1}, a publisher $P$ publishes two content objects $O_{j}$ and $O_{k}$. Further, two consumers $N_{A}$ and $N_{B}$ subscribe to access these protected contents $O_{j}$ and $O_{k}$. Furthermore, the object $O_{j}$ is published without encrypting, scrambling, or hashing the content name, while the object $O_{k}$ is published with encrypting, scrambling or hashing the content name to ensure its privacy. Assume that the consumer $N_{A}$ sent an interest packet $I_{A,j}$ encapsulating the access authorization information. In reply, based on the subscription information, if $N_{A}$ is a valid subscriber, the publisher $P$ encrypts the requested content segment $S_{0,j}$ with a consumer specific key and sends it to the consumer $N_{A}$. Based on the semantics of ICN and cache replacement schemes, the intermediate cache router $R_{5}$ stores the encrypted content segment $S_{0,j}$ for future use. Now lets suppose, the consumer $N_{B}$ requests the same content. If the meta-data of the encrypted stored packet is available to $R_{5}$, as in case of $O_{j}$, then the intermediate cache router $R_{5}$ will reply with the cached content $S_{0,j}$ to the consumer $N_{B}$. However, $N_{B}$ cannot decrypt the content segment $S_{0,j}$ as it was solely intended for $N_{A}$ and thus the payload is encrypted with the key known to $P$ and $N_{A}$. Contrarily, if the meta-data of the encrypted stored packet is unavailable to $R_{5}$, as in case of $O_{k}$, the interest packet $I_{B,k}$ will be forwarded to the publisher.


This issue can be solved by encrypting each content segment with a key known to all subscribers. In this regard, this issue can be viewed as a group key agreement problem. However, even in the presence of a perfect key distribution protocol, the assurance of backward and forward secrecy requires complex operations since the publisher in ICN has no control over the content after injecting it in the network. Moreover, in conventional group key agreement protocols~\cite{Bilal17Cluster,Zhang18}, the hosts share a cryptographic key for secure communications, which are not well-suited for the content centric ICN paradigm.

For example, if an authorized consumer unsubscribes from the service, then to ensure the forward secrecy it is necessary to make sure that leaving consumer don't have access to future keys for the group; hence, the shared key should be updated. From this point onward, the publisher would encrypt new version of content with updated group keys.  To access the content which is already disseminated in network caches, the authorized consumers need to keep both keys for effective cache utilization.  As shown in Figure~\ref{fig:1}(b), at time $t_{1}$, the publisher $P$ publishes object $O_{j}$ and shared the encryption key $K_{1}$ with all authorized consumers $N_{A}$, $N_{B}$ and $N_{C}$. Let's assume at time $t_{2}$ consumer $N_{B}$ unsubscribe with publisher $P$. The publisher $P$ will issue a new key $K_{2}$. Further assume that before unsubscribe event the copies of segment $S_{0,j}$ and $S_{1,j}$ were already disseminated in ICN core network; now if a consumer $N_{A}$ or $N_{C}$ request object $O_{j}$, it may get $S_{0,j}$ and $S_{1,j}$ from cache router encrypted with $K_{1}$ and rest of the segments from publishers encrypted with $K_{2}$. Similarly, if a new authorized consumer subscribes for the service, then to ensure the backward secrecy the shared key should be updated, and previous group members need to keep both keys for effective cache utilization.  Imagine a highly dynamic group where the consumers subscribe or unsubscribe very frequently, it will trigger numerous leave and join events, which will invoke group key agreement protocols each time. For effective caching, all consumers would keep record of multiple keys. Moreover, an extra decision operation is required to select a proper key; associating a time stamp can solve the problem at the cost of group member synchronization. Hence, the conventional group key management cannot handle the access control problem in ICN for ensuring the effective caching.

In our proposed scheme, we shifted the central target of keying process from hosts to data itself, i.e., the segments of the published content are encrypted with symmetric cryptographic keys that are unique to each segment and versions. The solution is to encrypt each content segment with a uniquely assigned key known to all subscribers; which raises three fundamental questions. How does one ensure that only an authenticated subscribed consumer can access the content? How can the consumer verify the originality of the content; that is, do we still need self-certifying? Finally, and most importantly, how can encryption keys be distributed among all of the consumers for each content segment? We answered all these questions in this work.

Specifically, we propose a secure distribution of protected content (SDPC) scheme, which consists of two protocol suites, 1) the keying protocol suite and 2) the subscription and content access protocol suite. The keying protocol suite enables the consumer and publisher to share a chain of secret keys required to decrypt the segments of the published content, while the subscription and content access protocol suite ensures that only authorized consumers receive the secret key generation information. 

The remainder of this paper is organized as follows. In Sections~\ref{sec:related} and~\ref{sec:SystemModel}, we summarize the related works and present the system model, respectively.  Section~\ref{sec:SDPC} describes SDPC with detailed discussions. Section~\ref{sec:SecurityAnalysis} presents an inclusive security analysis. In Section~\ref{sec:Performance}, we present the performance analysis of SDPC. Finally, we provide concluding remarks in Section~\ref{sec:Conclusion}.


\begin{figure*}
\centering
\includegraphics[width=1\textwidth]{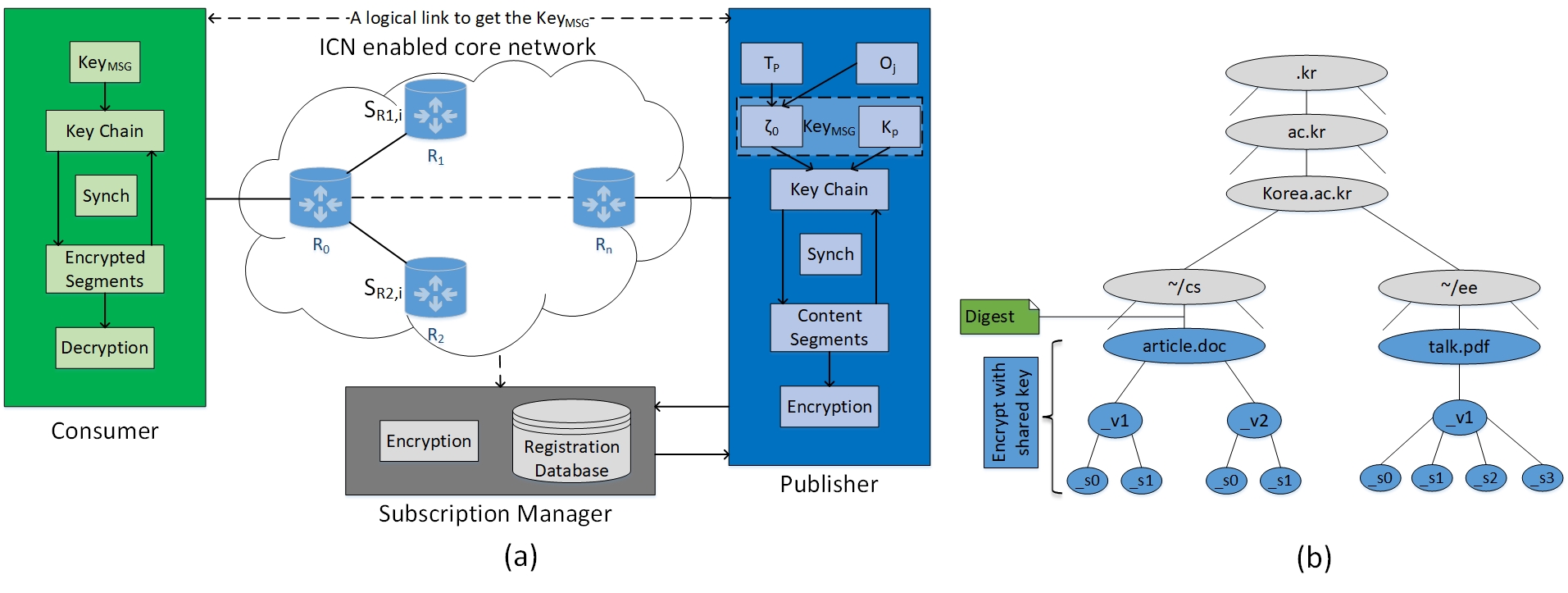}
\caption{Illustration of a) system model and b) naming scheme used in SDPC.}
\label{fig:2}       
\end{figure*}

\section{Related Works}
\label{sec:related}

Most existing access control schemes for secure contents are application specific or lack security strength. For example, in~\cite{Cho14}, the authors presented a scheme for protected contents using network coding as encryption. However, the scheme requires a private connection between the publisher and consumer to obtain the decoding matrix and missing data blocks. In~\cite{Xiaobin16}, the authors presented a security framework for the copyrighted video streaming in ICN based on linear random coding. It is proven that the linear random coding alone improves the performance of ICN~\cite{Bilal18System}. However in~\cite{Xiaobin16}, each video was encrypted with a large number of symmetric encryption keys, such that each video frame was encrypted with a unique symmetric encryption key. Since only authorized users who possessed the set of all keys could decrypt the video content, the distribution of a large number of keys for each video content can be an extra communication overhead.

 In earlier work~\cite{SMisra19}, the authors proposed a content access control scheme for ICN enabled wireless edge. The proposed one is an extension of ~\cite{SMisra13}, which employs the public-key based algorithm and shamir's secret sharing as a building block, named AccConF. To obtain a unique interpolating polynomial of shamir's scheme, AccConF espoused Lagrangian Interpolation technique. The calculation of Lagrangian Interpolation is a computationally expensive process. To reduce the client-side computational burden the publisher piggy backs an enabling block with each content, which encapsulates partially solved Lagrangian coefficients.

In work by ~\cite{Wood14}, an access control realized by a flexible secure content distribution architecture, which combins the proxy re-encryption and identity-based encryption mechanisms. The publisher generates a symmetric key and encrypt the content before dissemination. To access the content from in-network cache or directly from publisher, a consumer first sends a request to publisher to acquires the symmetric encryption key. Upon receiving the key request, the publisher validates and verifies the authenticity of consumer, and sends the symmetric key encapsulated in response message encrypted with consumer’s identity. The proposed scheme eliminated the asymmetric encryption, but it is not clear that how the consumer’s private identity could be known to the content provider. 

In other work ~\cite{Qzheng15}, author proposed a content access control scheme based on proxy re-encryption. In proxy re-encryption the content is re-encrypted by an intermediate node. In proposed scheme the edge routers perform the content re-encryption. Upon receiving a content request, the publisher encrypts the data and a randomly generated key k1, using its public key. Upon receiving the content request, edge router generates a random key k2 encrypted by the publisher’s public key and signed by the edge router. Edge router sends the encrypted k2 to publisher and appends the encrypted k2 with the content and dispatch it towards consumer. Meanwhile, the publisher verifies the authenticity of consumer, and generates the content decryption key K using K1, K2 and public key. Upon receiving K the consumer can decrypt the content. 

In other work~\cite{Qli17}, the authors proposed a distributed information flow control mechanism (named MCAC) to enable secure access control for the published content. In MCAC, the requests and content objects are labeled with \{$h$, $n$, $d$, $p$\}. These labels classify the contents based on the security and privacy requirements, where the h-level signifies the highest protection level and enforces non-caching policy, the n-level enforces the 1-level caching policy, the d-level permits multi-level caching policy, and the p-level supports all reading policy. To administer the MCAC information flow, the intermediate routers require to implement a trust computing base (TCB), consists of three modules; trust storage module (TSM), trust labeling module (TLM), and trust enforcement module (TEM). TSM governs the process of cryptographic session key establishment between participating routers and other nodes. The session keys are used to attain the h-level security by encrypting highly confidential $h$ labeled contents. TLM checks the label value and instructs the operating system accordingly to take further actions. TEM performs the content forwarding process and is responsible for content reclassification, i.e., TEM can re-labels a content to the h-level if it was at the n-level to hide the content based on privacy policy of the publisher. MCAC does not provide any mechanism to authenticate participating entities, which makes MCAC vulnerable to various attack. Moreover, to enforce the h-level security and privacy protection, all MCAC enabled routers need to establish a cryptographic session key and need to encrypt/decrypt all the communication between routers, which severely effects the performance of MCAC\footnote{ To verify the protocol claims, we implemented MCAC in an automated security protocol analysis tool, Scyther~\cite{Cremers12}, and also discussed its performance in Sections~\ref{sec:SecurityAnalysis} and~\ref{sec:Performance}.}.

In another study~\cite{Mannes15}, the authors presented an access control scheme for the encrypted content in ICN, which is based on the efficient unidirectional proxy re-encryption (EU-PRE) proposed by~\cite{Chow10}. The proposed scheme, named efficient unidirectional re-encryption (EU-RE), simplifies EU-PRE by eliminating the need of proxies in the re-encryption operation. However, the EU-RE scheme is still based on asymmetric cryptography, which is not suitable for several resource constraint applications such as, IoT and sensor networks. Moreover, the authors made an assumption that the content provider behaves correctly, i.e., it does not distribute any private content or decryption rights to unauthorized users. However, this assumption falsifies the protocol claims defined in~\cite{Cremers06}, which means EU-RE is weak against several attacks.\footnote{To verify the protocol claims, we implemented EU-RE in an automated security protocol analysis tool, Scyther~\cite{Cremers12}, and presented the results in Section~\ref{sec:SecurityAnalysis}.}

\section{System model}
\label{sec:SystemModel}

The system model used throughout this work is shown in Figure~\ref{fig:2}(a). For concrete discussion and better understanding, in the rest of the article, we present SDPC for a particular ICN architecture, i.e., named data networking (NDN)~\cite{Jacobson09}. However, SDPC can be adopted for other ICN architectures without changing the core idea.

In NDN core networks, we introduce a new entity, designated subscription manager $M$. The subscription manager $M$ can be a module installed on the publisher or it could be an independent entity in the network. In this work we assume that subscription manager $M$ is an independent entity associated with multiple publishers. We also assume that there is a secret number $n_{S}^{i}$ associated with each valid subscriber (or consumer) $i$, which is registered with the subscription manager $M$. The registration could be made offline or online using a smart mechanism. The subscription manger $M$ stores the secret number $n_{S}^{i}$ in a hash table, which is a part of registration database, as shown in Figure~\ref{fig:2}(a). Note that being registered does not mean the consumer is entitled to access a certain protected content.  When a registered consumer is interested in a protected content, the consumer should first subscribe to the protected content, for instance, subscribing to a movie channel. In the first step, the consumer sends an interest request for the protected content along with the subscription request, and the publisher routes the request towards the subscription manager $M$. After that, the subscription manager $M$ authenticates the consumer $N_{A}$ and in response the publisher $P$ sends the encryption key generation information $KEY_{MSG}$. Using $KEY_{MSG}$ as a seed for a secure hash function, the consumer $N_{A}$ and the publisher $P$ can generate a chain of keys. Then, the publisher $P$ uses these keys to encrypt the segments of the published content; likewise, after acquiring $KEY_{MSG}$ the consumer $N_{A}$ generates the same keys to decrypt the segments of the published content. 

To acquire $KEY_{MSG}$, the first interest packet sent by a consumer should reach the publisher. To avoid any cache hit, it is important the name of the content should be unique between the consumer and the publisher, yet it should identify the requested object. As shown in Figure~\ref{fig:2}(b), the name of the segment $0$, "korea.ac.kr/$\sim$fil/test.doc/\_v1/\_s0", is a variable length and in a human readable form. However, to avoid any cache hit the SDPC naming scheme adopts the name uniqueness, such that the name of first interest is unique between consumer and publisher. To, achieve the name uniqueness the consumer inserts the hash of the secret number $n_{S}^{i}$, and encrypts the content name with $K_{TS}^{i}$. Then, the name of the segment $0$ becomes "korea.ac.kr/$\sim$fil/$Hash \left( n_{S}^{i} \right)$/$E_{TS}^{i} \left( test.doc/ \_v1/ \_s0 \right)$". In this naming scheme the insertion of digest and encryption of naming part provides a consumer-publisher specific unique name and as a result the interest packet always reaches the publisher without any cache hit. Note that the usage of consumer name space is restricted for acquiring $KEY_{MSG}$ only, this gives provides prevention against DoS attacks.

After acquiring $KEY_{MSG}$ the consumer can access the rest of the segments by using a shared authoritative name space. The name for each segment includes a hash digest $Hash \left( KEY_{MSG} \right)$, and the object name is encrypted with a uniquely assigned key $K_{l}^{j}$, which is generated using $KEY_{MSG}$ for each segment $l$ of an object $O_{j}$. For example, the name for segment $\_ s1$ of object $O_{j}$ is given by "korea.ac.kr/$\sim$ fil/$ Hash \left( KEY_{MSG} \right)$/$E_{1}^{j} \left( test.doc/\_v1/ \_s1 \right)$". With the insertion of $Hash \left( KEY_{MSG} \right)$ and encryption of the naming part with keys generated using $KEY_{MSG}$,\ this naming scheme provides a shared authoritative name space for all authorized consumers and thus it enables an effective content caching.  Moreover, this naming scheme ensures the privacy, because the content name is invisible to outsider without any knowledge of $n_{S}^{i}$, $KEY_{MSG}$, and cryptographic keys $K_{TS}^{i}$ and $K_{l}^{j}$.

Let's suppose Figure~\ref{fig:2}(a), a consumer sends an interest packet $I_{A,i}$ utilizing the proposed naming scheme. Then, the packet will reach publisher without any cache hit. Let us say that protected content object $O_{i}$ is composed of $k$ segments of $ S= \{ S_{1,i}, S_{2,i} \ldots S_{k,i} \}$; further, the intermediate cache routers $R_{1}$ and $R_{2}$ have the copies of the protected content segments, represented by $S_{R1,i} \subseteq S$  and $S_{R2,i} \subseteq S$. If the consumer is a valid subscriber, the publisher sends the encryption key generation information $KEY_{MSG}$ to the consumer. After receiving the key generation information, the consumer can decrypt the content segments, which may be delivered directly from the intermediate cache router.

\section{Secure Distribution of Protected Content}
\label{sec:SDPC}

SDPC consists of two protocol suites: 1) the keying protocol suite and 2) the subscription and content access protocol suite. The keying protocol suite is comprised of a key generation protocol and a key agreement protocol for content protection. Likewise, the subscription and content access protocol further includes three protocols, one dealing with the consumer subscription and the other two dealing with access to the protected contents published by different publishers. 

\subsection{Keying Protocol Suite}

In the keying protocol suite, the key generation protocol generates a commitment key using an irreversible function similar to the ones used in~\cite{Bilal17JCS} \cite{Bilal17patent}. The commitment key is further used to drive multiple keys; for instance, a chain of content segment encryption keys, a ticket encryption key, and a consumer associated symmetric key are derived from the commitment key.

\begin{figure*}
\centering
\includegraphics[width=0.82 \paperwidth]{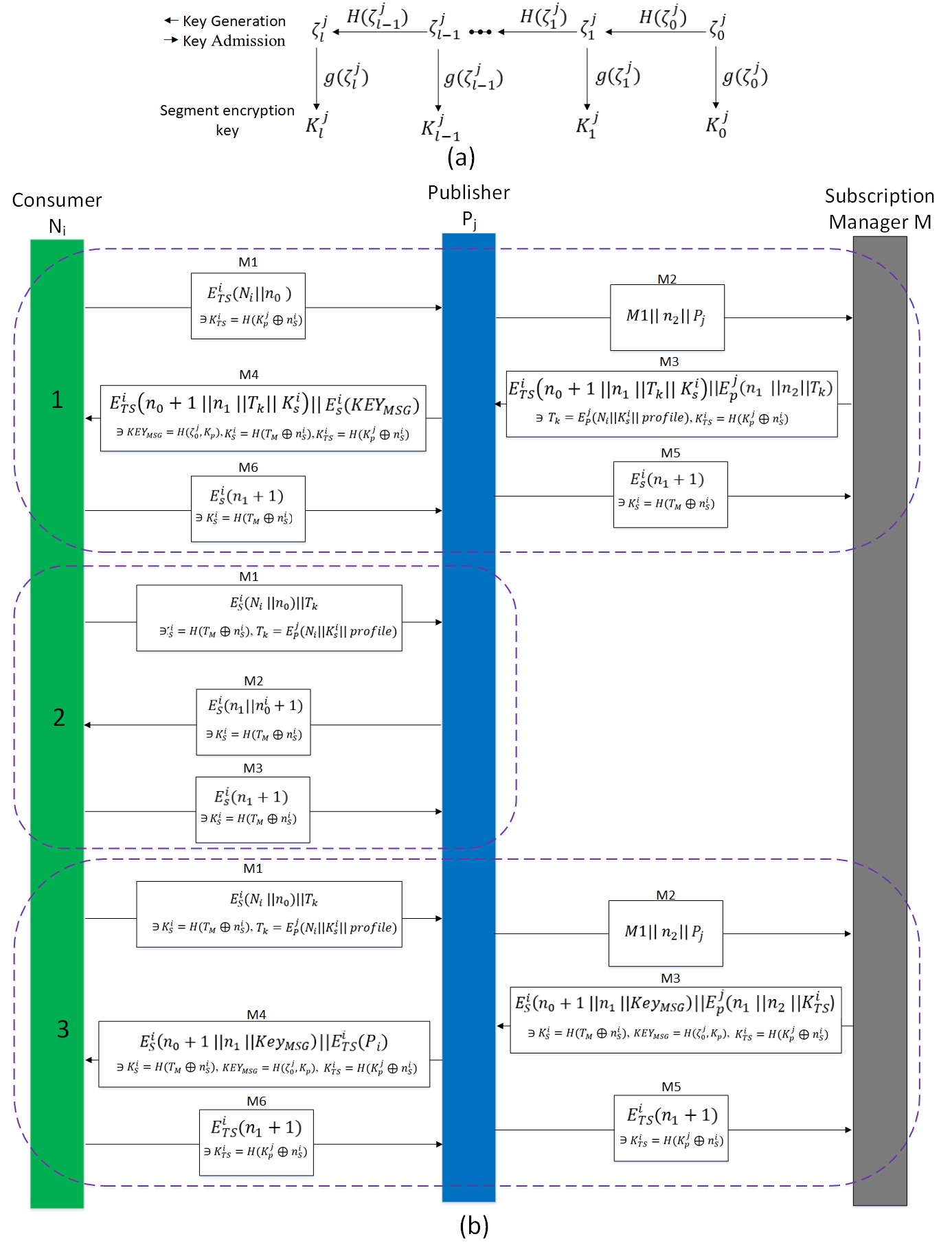}
\caption{The SDPC protocol suite: (a) symmetric keys generation and admission with reference to segment number of protected content and (b) message exchange for SubP, APSub, and APSub3.}
\label{fig:3}       
\end{figure*}

The key generation mechanism for the content protection is shown in Figure~\ref{fig:3}(a). First, the publisher divides a large content into equal sized segments. For each protected content object $O_{j}$, the publisher generates a unique commitment key generator by using an irreversible one-way hash function $\zeta_{0}^{j}=H \left( T_{P}, O_{j} \right)$, where $T_{P}$ is the time of publishing and $O_{j}$ represents the content name and version\footnote{Each version of content object is encrypted with a separate chain of keys. It empowers the publisher to control version-based access.}. After that, the publisher generates a chain of key generators of the length $L=\frac{sizeof \left( O_{j} \right) }{segment size}$ by using an irreversible one-way function $\{ H \left(  \zeta _{0}^{j} \right) = \zeta _{1}^{j}, H \left(  \zeta _{1}^{j} \right) =\zeta _{2}^{j}, \ldots, H \left(  \zeta _{l-1}^{j} \right) = \zeta _{l}^{j} \}$. Each generator $\zeta$ in the chain is used by a function $g$ at a specific index location in the chain to derive a content segment encryption key. For instance, at index $k$, the function $g \left(  \zeta _{k}^{j} \right) =H \left(  \zeta _{k}^{j},K_{p} \right)$ generates the key $K_{k}^{j}$ used for encrypting the $k$th segment of the content object $O_{j}$, where $K_{p}$ is the public key of the publisher. The use of $K_{p}$, in symmetric key generation process, implicitly ensures the originality of the content, i.e., the content are still self-certifying with out use of expensive asymmetric encryption. For instance, very efficient public key algorithms, such as ECC~\cite{koyama91}, are almost three thousand times slower than symmetric key algorithms~\cite{Wong06} such as RC5~\cite{Rivest94}. The symmetric keys generated as a result of the SDPC keying protocol have the size of 256 bits. Hence, in the subsequent section on the authentication protocols, any symmetric encryption algorithms supporting the 256-bit key can be used, e.g., RC5/6~\cite{Rivest94}, Rijndael~\cite{DaemenAES}, and Twofish~\cite{SchneierTwo}.

\subsection{Subscription and Content Access Protocol suite}
When a consumer wants to subscribe to the protected content, the consumer gains an initial access using a subscription protocol (SubP). After SubP, the consumer can use a ticket to access multiple protected contents published by the publishers or managed by a third party. 


\subsubsection{Initial Access and Subscription Protocol (SubP)}

If a consumer $N_{i}$ wants to subscribe to the protected content (e.g., subscribing for a movie channel), $N_{i}$ first generates an encryption key $K_{TS}^i=H(K_p^j\oplus n_S^i)$, where $K_{p}^{j}$ is the public key of the publisher and $n_{S}^{i}$ is a secret number shared with the subscription manager $M$. SubP continues as follows:

\begin{enumerate}
	\item  As shown in \textbf{1} at Figure~\ref{fig:3}(b), $N_{i}$ injects a subscription interest packet  $I_{i}$ and the NDN core network forwards it to the publisher $P_{j}$. The interest packet encloses $n_{0}^{i}$ that is encrypted with the generated encryption key $K_{TS}^{i}$. 

	\item  Upon receiving the request from $N_{i}$, $P_{j}$ forwards the request in conjunction with its identity and the challenge $n_{2}$ to the subscription manager $M$. Note that $P_{j}$ cannot decrypt the part of the interest packet which is encrypted with key $K_{TS}^{i}$ and registration number $n_{S}^{i}$ remains invisible to the publisher.

	\item  $M$ retrieves the profile of $N_{i}$ from the database. If $N_{i}$ is a legitimate consumer, $M$ generates the keys $K_{TS}^i=H(K_p^j\oplus n_S^i)$ and $K_{S}^{i}=H \left( T_{M}\oplus n_{S}^{i} \right)$, and sends $u_{0}=E_{TS}^{i} \left( n_{0}+1 \vert  \vert n_{1} \vert  \vert T_{k} \vert  \vert K_{s}^{i} \right)$ to $P_{j}$, where $T_{M}$ is the time of issuing the session key $K_{S}^{i}$. The message M3 includes a ticket $T_{k}=E_{P}^{j} \left( N_{i} \vert  \vert K_{s}^{i} \vert  \vert  profile \right)$, a challenge $n_{1}$ for $N_{i}$, and a challenge response for $P_{j}$. After that, $ P_{j}$ verifies the challenge response and stores $n_{1}$ to use it as a message authentication in M5 and M6. In addition, $P_{j}$ retrieves the profile and the session key $K_{s}^{i}$ from the ticket. Since ticket is encrypted with the public key of $P_{j}$, the consumer  $N_{i}$ cannot decrypt it, but can use it to subscribe to other contents published by $P_{j}$, without contacting the subscription manager $M$.
	
	\item  $P_{j}$ forwards $u_{0}$ to $N_{i}$ along with $ KEY_{MSG}=H \left(  \zeta _{0}^{j},K_{p} \right)$, which is required to decrypt the segments of the published content and also used as a content object identifier. After verification of a challenge $\left( n_{0}+1 \right)$, $N_{i}$ accepts $T_{k}$ and generates a key chain to decrypt the protected content. The generated key chain involves the public key of $P_{j}$, hence, the content is also self-certifying.

	\item  $P_{j}$ sends the challenge response $\left( n_{1}+1 \right)$ to $M$ for the confirmation of a successful protocol run. After challenge $\left( n_{1}+1 \right)$ confirmation, $P_{j}$ may optionally register $N_{i}$ in its own database. If $P_{j}$ does not receive any challenge response within a certain period of time, $P_{j}$ marks $T_{k}$ as a stolen ticket.
\end{enumerate}

In SubP, secure exchanges of $n_{0}$, $n_{1}$, and $n_{2}$ ensure the message authentication between the consumer and the subscription manger, between the subscription manger and the publisher, and between the publisher and subscription manger, respectively. Likewise, the message authentication between the consumer and publisher is established by the session key encryption and $n_{1}$.

\subsection{Content Access Protocols}
\subsubsection{Access Protocol after Subscription (APSub) }

When the consumer $N_{i}$ wishes to access some other protected contents published by the publisher $P_{j}$, $N_{i}$ sends an interest request for the protected content along with the ticket $T_{k}$ and APSub continues as follows.

\begin{enumerate}
	\item  As shown in \textbf{2} at Figure~\ref{fig:3}(b), $N_{i}$ injects a subscription interest packet, enclosing $Access_{req}=E_{S}^{i} \left( N_{i} \vert  \vert n_{0} \right)  \vert  \vert T_{k}$. The NDN core network forwards it to the publisher $P_{j}$. The publisher $P_{j}$ decrypts the ticket and verifies the sender's identity $N_{i}$  by retrieving $K_{s}^{i}$. If the value $N_{i}$ does not match, $P_{j}$  will ignore the request and otherwise proceed as follows.

	\item  $P_{j}$ sends a challenge response along with the new challenge encrypted with the session key $K_{s}^{i}$. $P_{j}$ also send $KEY_{MSG}$, which is required to decrypt the segments of the published content.

	\item  $N_{i}$ sends a challenge response $n_{1}$. If $P_{j}$ does not receive the challenge response within a certain period of time, $P_{j}$ marks $T_{k}$ as a stolen ticket. 
\end{enumerate}

In APSub, the secure exchange of $n_{0}^{i}$ ensures the message authentication between the consumer and the publisher.

\subsubsection{Access Protocol after Subscription involving a Third Party (APSub3)}

Assume a consumer $N_{i}$ subscribed with $P_{i}$, which means it shares a session key $K_{s}^{i}$ with $P_{i}$ and holds a $T_{k}$  encrypted with public key of $P_{i}$. Now if $N_{i}$ wishes to access the protected contents published by a third-party content publisher $P_{j}$, APSub3 continues as follows.

\begin{enumerate}
	\item  As depicted in \textbf{3} at Figure~\ref{fig:3}(b), $N_{i}$ injects a subscription interest packet enclosing $Access_{req}=E_{S}^{i} \left( N_{i} \vert  \vert n_{0} \right) \vert  \vert T_{k}$ and the packet is forwarded to the publisher $P_{j}$.

	\item  Upon receiving the request from $N_{i}$, $P_{j}$ forwards the request in conjunction with its identity and the challenge $n_{2}$ to $M$. Note that $P_{j}$ cannot decrypt  $Access_{req}$ in the interest packet that is encrypted with the key $K_{s}^{i}$, which a shared session key between $N_{i}$ and $P_{i}$, which ensures the third-party content distributor cannot misuse the consumer secure information, such as profile and secret share number.

	\item $M$ retrieves the profile from $T_{k}$, and if $N_{i}$ is a legitimate consumer, $M$ generates the key $K_{TS}^i=H(K_p^j\oplus n_0)$, and sends $u_{0}=E_{S}^{i} \left( n_{0}+1 \vert  \vert n_{1} \vert  \vert KEY_{MSG} \right)$  to  $P_{j}$. The message M3 also includes the key  $K_{TS}^{i}$, a challenge $n_{1}$ for $N_{i}$, and the challenge response for $P_{j}$, which are encrypted with the public key $P_{j}$. After that, the publisher $P_{i}$ verifies the challenge response and stores $n_{1}$. Note that the ticket is encrypted with the public key of $P_{i}$. Therefore, $N_{i}$ and third-party publisher $P_{j}$ cannot decrypt it. Also, $ KEY_{MSG}$ is inaccessible to $P_{i}$, which ensures that the third-party content distributor cannot misuse the protected content.

	\item  $P_{j}$ forwards $u_{0} \vert  \vert E_{TS}^{i} \left( P_{j} \right)$  to $N_{i}$. After the verification of the challenge $\left( n_{0}+1 \right)$, $N_{i}$ generates $K_{TS}^i=H(K_p^j\oplus n_0)$  and sends the challenge response $\left( n_{1}+1 \right)$ to $P_{j}$. Now $ N_{i}$ can generate a key chain to decrypt the protected published content. Since the key chain is generated using the public key of $P_{i}$, the content is also self-certifying.

	\item  $P_{j}$ sends the challenge response $\left( n_{1}+1 \right)$ to $M$ for the confirmation of a successful protocol run.

	\item  After the challenge confirmation, $P_{j}$ closes the protocol run. If $P_{j}$ does not receive any challenge response within a certain period of time, $P_{j}$ marks $T_{k}$ as a stolen ticket.  
\end{enumerate}

In SubP3, secure exchanges of $n_{0}$, $n_{1}$, and $n_{2}$ ensure the message authentication between the consumer and the subscription manger, between the subscription manger and the third-party publisher, and between the third-party publisher and subscription manger, respectively. Likewise, the message authentication between the consumer and the third-party publisher is established by a temporary session key $K_{TS}^{i}$ and $n_{1}$.

\section{Security Analysis}
\label{sec:SecurityAnalysis}

This section presents an inclusive security analysis, formal analysis using BAN logic~\cite{Burrows90}, and Scyther implementation results~\cite{Cremers12}. 

\subsection{Naming based Attacks}

In NDN the objects are identified by a human readable naming system, which can lead to watchlist and sniffing attacks~\cite{Jakobsson17,Abdallah17,Klump10}. 

In watchlist, an attacker who has control over communication links and cache routers, can delete or filter the content based on a predefined list of content objects. With the use of SDPC, the content is encrypted and invisible to the attacker. Recall that in NDN it is not obligation that a content object must carry an explicit content name, rather it can carry an implicit content identifier computed from the corresponding interest. This solution hides the object from the attacker. Let us reconsider the example in Figure~\ref{fig:2}(b). The first interest packet carries the name "korea.ac.kr/$\sim$fil/ \(Hash( \left( n_{S}^{i} \right) \) / \( E_{TS}^{i} \left( test.doc/ \_ v1/ \_ s0  \right)) \)"  beside the insertion of hash digest \(hash( \left( n_{S}^{i} \right) \),  the object name is encrypted with  \( K_{TS}^{i} \) .\  After acquiring  \( KEY_{MSG} \)  the name for \_s1 is then given by "korea.ac.kr/$\sim$fil/ \(  Hash \left( KEY_{MSG} \right)  \)  / \(  E_{1}^{j} \left( test.doc/ \_ v1/ \_ s1  \right)  \)". The attacker cannot get $KEY_{MSG}$, $n_{S}^{i}$, $ K_{TS}^{i}$, and $K_{l}^{j}$, and thus launching watchlist attack is not possible. Moreover, it completely hides the object name from the attacker, which ensures the privacy of the consumer.

Contrarily, in a sniffing attack the intruder does not have any list of pre-defined contents, rather it monitors the network and filters or eliminates the data if it contains some specified keywords. Such sniffing attack is not possible in SDPC, because the data is encrypted with the secret keys.

\subsection{DDoS Attacks}
The in-network caching makes NDN intrinsically resilient against distributed denial of service (DDoS) attacks~\cite{Wang18,Xin16}.  DDoS is a malicious attempt to disturb normal traffic to a server, for instance, multiple compromised systems send fake interest packets to a content publisher. Once the content is disseminated across network caches, the DDoS attack against a publisher depletes due to the on-path cache hits. However, assume somehow an attacker manages to flood all interests to a targeted publisher. With the use of SDPC, the total burden after a successful DDoS attack on the targeted publisher will remain insignificant.  This is because the subscription manager $M$ keeps the record of registered nodes in the hash table, which entries represent $K_{TS}^i=H(K_p^j\oplus n_0)$ session keys. Thus, in case of suspicion, the subscription manger in SDPC can identify fake requests by a hash table lookup with the complexity $O \left( 1 \right)$.

\subsection{Time Analysis Attack}
In NDN any cache node can store content segments. An intruder can guess that a particular content was requested by a user in particular vicinity by observing the request response time of a cached or uncached content. With the use of SDPC the payload is encrypted with one of the key derived from $KEY_{MSG}$ and the name of an object is identified by the digest and encrypted fields.  Since the intruder cannot acquire $KEY_{MSG}$ on time, it cannot create a valid request to launch a time analysis attack.

\subsection{Unauthorized Access}
SDPC allows the caches to store encrypted contents and to use a naming scheme unrecognizable to intruders. An intruder can access the content only after acquiring $KEY_{MSG}$. Since the delivery of $KEY_{MSG}$ in SDPC is achieved by handshake messages, where each message exchange contains an explicate (nonce challenge) or implicit (encryption key derived from nonce) message authentication; further each message is encrypted with $K_{TS}^i=H(K_p^j\oplus n_0)$, for unauthorized access an intruder needs to acquire $K_{TS}^i$.

\subsection{Traffic Monitoring Attack}
In traffic monitoring attack \cite{Tlau}, an intruder targets a consumer and tries to identity the requested contents. To launch a traffic monitoring attack the intruder takes control of edge router and observes all the requests send by the target consumer. However, in SDPC the content name is encrypted, which hides the object name from the attacker, consequently the traffic monitoring cannot reveal the identity of requested contents.

\subsection{Formal Analysis using BAN Logic}

BAN logic~\cite{Burrows90} is widely used for the formal analysis of security protocols, till recently\cite{frash16} \cite{Mahood18} . To verify the security of the SDPC protocol suite, it is sufficient to demonstrate the security of SubP since the rest protocols are extensions of SubP. The BAN logic analysis shows that SDPC is safe against large number of attacks. A detailed formal analysis of SDPC using BAN logic can be found in Appendix-I or at~\cite{SDPC_App} .

\subsection{Scyther Implementation Results}

Although BAN logic provides a foundation for the formal analysis of security protocols, a few attacks can be undetectable even with BAN logic~\cite{Nessett90}. However, the critical analysis of BAN-logic in~\cite{Nessett90} is based on usage of asymmetric cryptography, whereas SDPC utilizes symmetric cryptography. Furthermore, \cite{Nessett90} argue that BAN-logic methodology is faulty because it is assumed that physical security and administration do not suffer from the loss of messages by the underlying communication facility or because of host crashes. Owing to replication of contents across the network, in ICN this assumption has minor effect. Still, for the additional proof of the strength of the SDPC protocol suite, we implemented SDPC, EU-RE~\cite{Mannes15}, and MCAC~\cite{Qli17} in an automated security protocol analysis tool, Scyther~\cite{Cremers12}.

We considered four claims: 1) aliveness, 2) weak agreement, 3) non-injective agreement, and 4) non-injective synchronization~\cite{Cremers06}. These four claims are proven to be true for SDPC  by using BAN logic. In Scyther, a protocol is modeled as an exchange of messages among different participating roles.  For instance, in NDN-SDPC, the consumer and publisher are in the roles of initiator (I) and responder (R), respectively, whereas the subscription manger is in the role of a server (S). In EU-RE, the publisher acts both as a responder and  as a server (R{\_}S), whereas in MCAC the consumer and publisher are in the roles of I and R, respectively, whereas the 3rd party authenticator has the role of S. The Scyther tool integrates the authentication properties into the protocol specification as a claim event. We tested SDPC, MCAC, and EU-RE by employing the claims mentioned earlier, with the parameter settings given in Table~\ref{tab:3}.

\begin{table}
\centering
\caption{Scyther tool parameter settings.}
\label{tab:3}       
\begin{tabular}{c|c}\hline{\smallskip}
Parameter  &	Settings   \\ \hline \hline
Number of runs  & 1 to 3\\ \hline
Matching type  & Find all types of flaws\\ \hline
Search pruning  & Find all attacks\\ \hline
Number of patterns per claim  & 10\\ \hline
\noalign{\smallskip}           
\end{tabular}
\end{table}

\begin{table*}[ht]
\caption{Scyther results for SDPC, MCAC, and EU-RE.}\label{tab:4}
\centering
\begin{tabular}{c|ccccccccccc}
\hline{\smallskip}
\multirow{1}{*}{{\bf Claims}} & \multicolumn{3}{C{1.5cm}}{\bf MCAC \cite{Qli17}}& \multicolumn{3}{C{1.5cm}}{\bf MCAC \cite{Qli17} auth.}& \multicolumn{2}{C{1.5cm}}{\bf EU-RE \cite{Mannes15}}& \multicolumn{3}{C{1.5cm}}{\bf NDN-SDPC }  \\ \hline \hline

& $I$ & $R$ & $S$ & $I$ & $R$ & $S$ & $I$ & $R_S$ & $I$ & $R$ & $S$  \\ \hline
Aliveness & N & N & N & Y & Y & Y & Y & Y & Y & Y & Y  \\ \hline
Weak Agreement & N & N & N & Y & Y & Y & N & Y & Y & Y & Y \\ \hline
Non-injective Agreement & N & N & N & Y & Y & Y & N & Y & Y & Y & Y \\ \hline
Non-injective Synchronization & N & N & N & Y & Y & Y & N & Y & Y & Y & Y \\ \hline 
\end{tabular}
\begin{tabular}{@{}cc@{}}
\multicolumn{1}{p{\linewidth-2cm}}{\footnotesize \center{N = Protocol claim is not fulfilled; Y = Protocol claim is fulfilled}}
\end{tabular}
\end{table*}

The Scyther results are shown in Table~\ref{tab:4}. It is clear that SubP qualifies all of the protocol claims and no attacks were found. Consequently, for a large number of systems and scenarios, SDPC guarantees safety against a large number of known attacks such as impersonating, man-in-middle, and replay attacks. However, in EU-RE, the author made an assumption that the content provider behaves correctly, i.e., it does not distribute any private contents or decryption rights to unauthorized users, this assumption falsifies the protocol claims, which means EU-RE is weak against several attacks. The Scyther implementation shows that initiator fails to confirm claims 2, 3, and 4.

In MCAC, the TCB along with encrypted communication between routers provide strong security against man-in-middle attack; however, during the bootstrapping, the session key establishment is conducted by the Diffie-Hellman (DH) key distribution algorithm ~\cite{Merkle1978} without using a proper authentication procedure. Since the DH algorithm does not inherently provide authentication, it can be secure only if it is properly integrated with another authentication protocol. This weak link in MCAC makes it vulnerable to several attacks, even a man-in-middle attack could be possible if an intruder tempered the session key distribution procedure during bootstrapping process. Therefore, from Scyther implementation results, it can be seen that MCAC fails to qualify a signal claim; further, if we assume the DH key exchange protocol is integrated with an authentication protocol or bootstrap process is hidden from the intruder, then MCAC qualifies all the claims. The inclusion of the authentication process causes the extra processing burden only during the bootstrap process and can be ignored for the next steps in the protocol.

\section{Performance Evaluation}
\label{sec:Performance}

\begin{figure}
\centering
\includegraphics[width=0.5\textwidth]{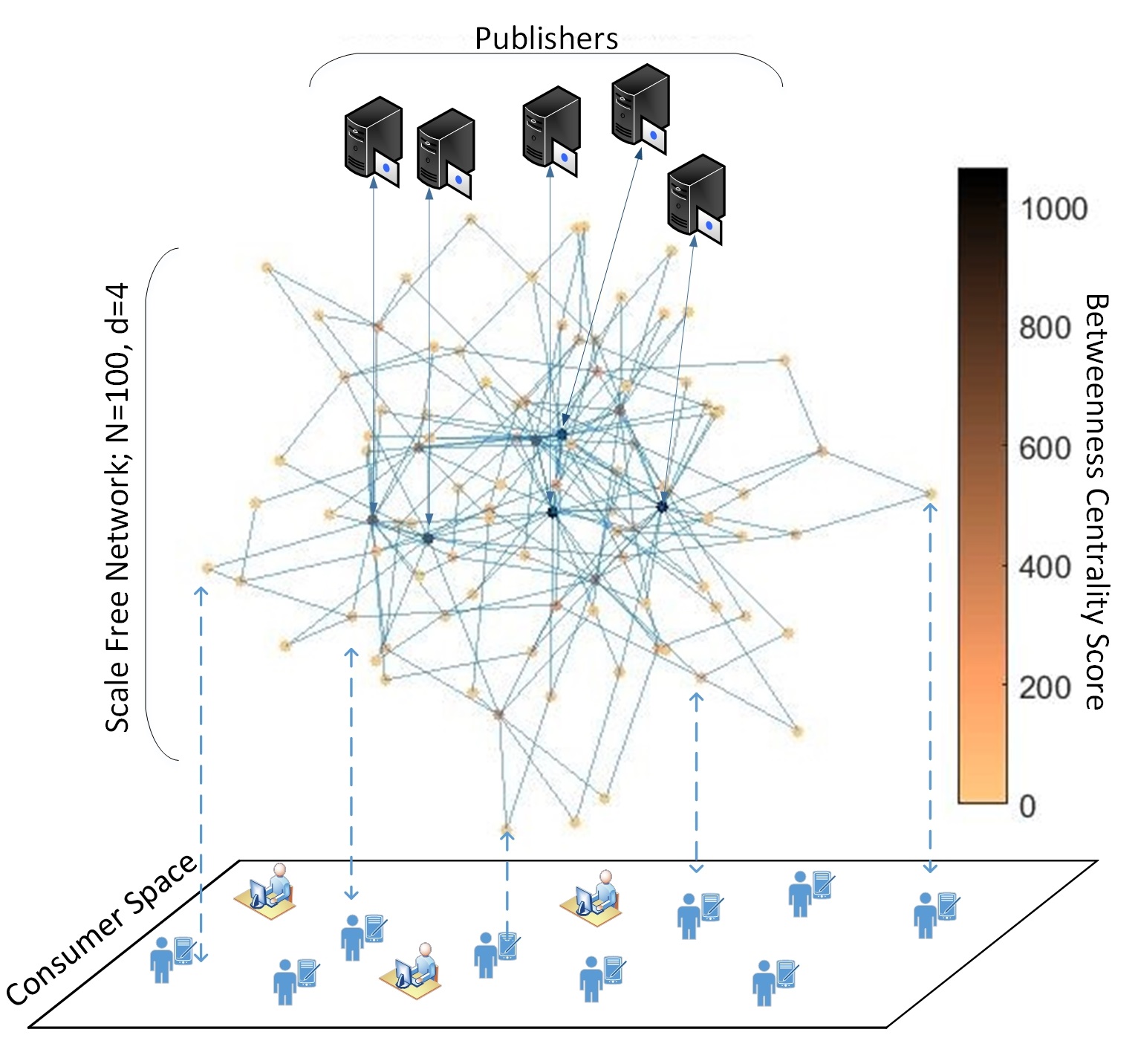}
\caption{Network setup for performance evaluation.}
\label{fig:4} 
\end{figure}

We consider a scale-free network of 200 cache nodes generated using the Barabási–Albert (BA) model, as shown in Figure~\ref{fig:4}, which connects the publisher and the consumer space. Each cache router has a static request routing table. Further, we assume five content publishers in the network. Each publisher has 100,000 secure content items, and a Zipf-distribution with a popularity distribution exponent $\alpha=0.7$ is used to determine the population of content items in the entire network. 

To ensure quick dissemination of the contents in the network, the publishers are connected to the cache routers with the highest betweenness centrality score, it helps to bringing system in steady state in short time. Furthermore, 25 gateway cache routers are connected to the consumer space with a large number of consumers. At any given time, 400~500 consumers are subscribed with each publisher and thus the total number of consumers subscribed to five publishers varies between 2,000 and 2,500. The size of each content item is 1GB, which is divided into 10 segments, and the link capacity between two cache routers is 1 Gbps, Finally, least frequently and recently used (LFRU)~\cite{Bilal17Access} is used in the experiment as a content replacement scheme.

We implemented the network setup, as described above, in MATLAB and compared the performance of NDN-SDPC against MCAC, EU-RE, and native NDN for two scenarios, 1) using end-to-end encryption, which makes the caching ineffective, as discussed in Figure~\ref{fig:1}, and 2) enabling the caching with a conventional way of a shared group key~\cite{Bilal17Cluster}\cite{Zhang18}. In the scenario 2, the shared group key enables in-network caching, but the shared group key is unfeasible because the authorized consumers need to keep a large number of keys for effective cache utilization in highly dynamic environments. Moreover, extra decision operations are required to select a proper key and to determine the timing of key deletion. For simplicity, in the scenario 2, we only consider the computational and message complexity required to ensure backward and forward secrecy. The processing required to select an appropriate key on the consumer side is ignored. Further, the scenario 2 is simulated for different levels of dynamicity in the consumer space, by considering 5, 15, and 25 leave and join requests per unit time; representing cases 1, 2, and 3, respectively. This comparison is made in terms of average download time\footnote{The average download time is defined as the ratio of the total number of requests observed on all 25 cache routers to the time taken to receive all the requested contents at the gateway routers.}, publisher load\footnote{The publisher load is defined as the percentage of interests reached at publisher. High the publisher load implies low cache hit.}, and timeout interest ratio\footnote{The timeout interest ratio is defined as the percentage of interests timed out and re-transmitted.}.

\begin{figure}
\centering
\includegraphics[width=0.5\textwidth]{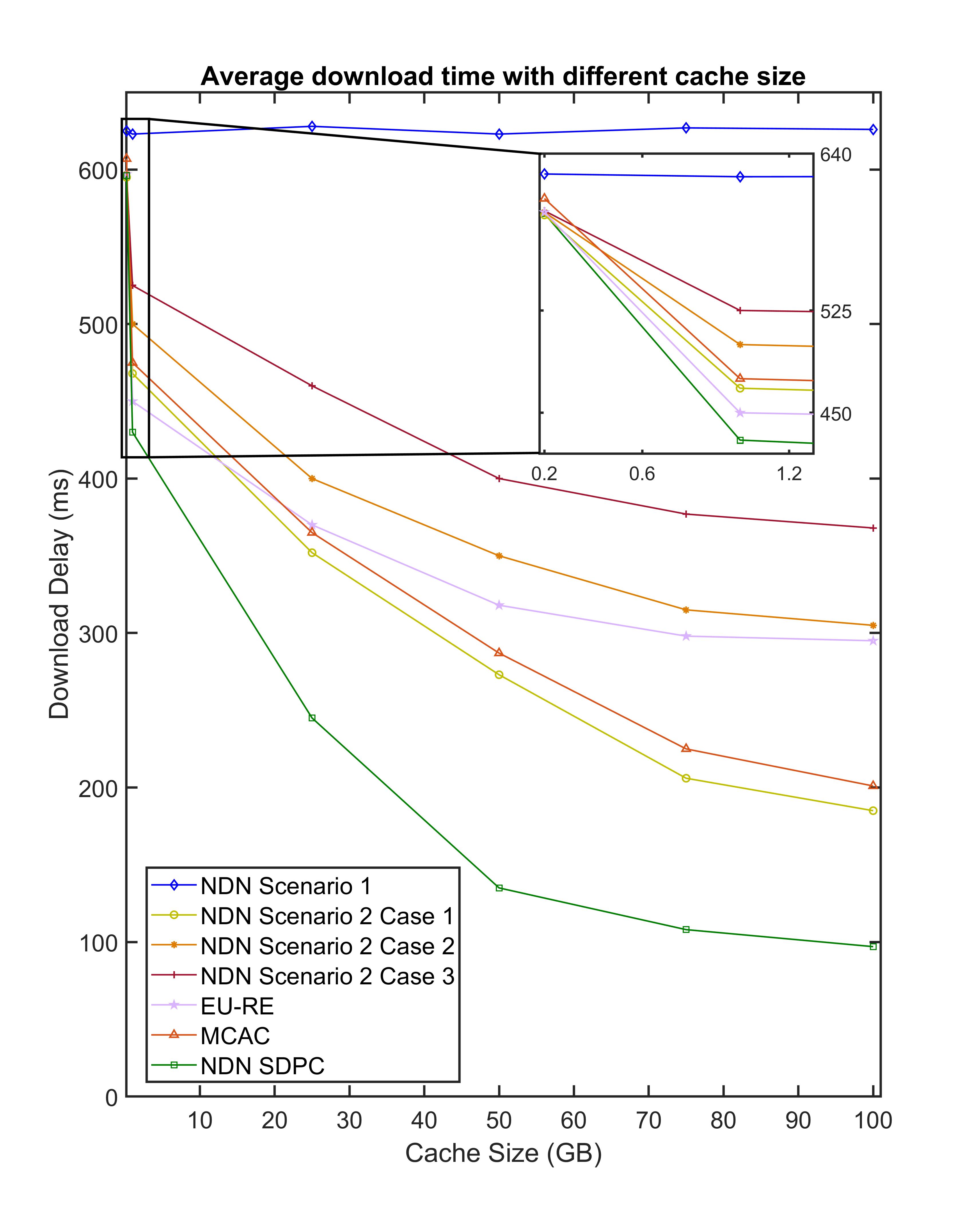}
\caption{Average download delay at gateway-cache router.}
\label{fig:5} 
\end{figure}

\begin{figure}
\centering
\includegraphics[width=0.5\textwidth]{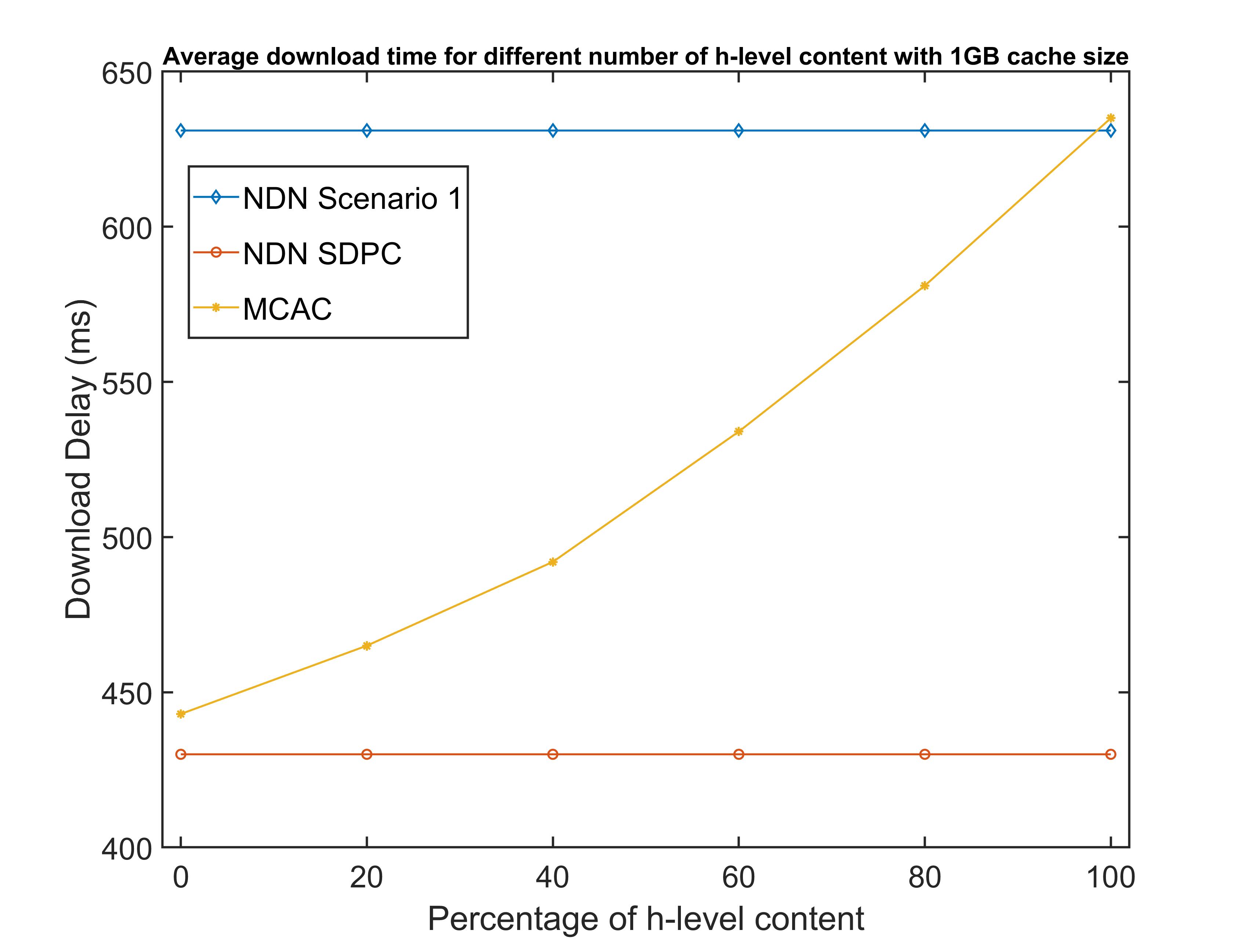}
\caption{ Average download delay for different numbers of h-level contents.}
\label{fig:6} 
\end{figure}

Figure~\ref{fig:5} shows the comparison of the average download time observed when each of the 25 gateway cache routers receives the requests that are generated by a Poisson distribution with a rate $\lambda_j^d=100$ req/s.  The results are considered for different cache sizes of 200 MB to 100 GB; further, in case of MCAC, it is considered that 20\% of contents are labelled as the h-level and 80\% as the d-level.\footnote{The existence of different levels of content impacts the overall performance of MCAC, as shown in Figure~\ref{fig:6}.} From Figure~\ref{fig:5}, it can be seen that NDN with SDPC outperforms  EU-RE and native NDN both in the scenarios 1 and 2. The performance of NDN in the scenario 2 degrades further with the increase in dynamicity of the consumer space. The performance results of EU-RE are interesting, for smaller cache size [200MB-1GB], the performance of EU-RE is very close to NDN-SDPC, and it outperforms native NDN both in the scenarios 1 and 2. However, the performance gap increases with the increase of the cache size, and it falls down below NDN in the scenario 2 with case 3. In Eu-RE, the key revocation and content version are not correlated, and this can be one of the reason of such performance degradation. For a large cache size, MCAC performs better than EU-RE and NDN in scenario 2 with cases 2 and 3; however, NDN-SDPC and NDN in scenario 2 with case 1 outperform the MCAC. As discussed earlier, MCAC enforces intermediate routers to implement TCB, which includes several operations and encryption/decryption process for the h-level and the n-level secure content, these extra operations introduce processing delay at intermediate routers, the performance of MCAC further decreases with increasing the number of h-level contents.

Figure~\ref{fig:6} shows the comparison of average download time comparison between NDN, NDN-SDPC and MCAC, for different numbers of h-level contents ranging from 0 to 100\% of total traffic, with the fixed 1GB cache size. From Figure~\ref{fig:6}, it is clear that performance of MCAC degrades with increasing number of h-level content.  The performance degradation of MCAC with increasing the number of the h-level contents is quite obvious, because h-level contents require no caching policy; hence, all interest packets traverse to the publisher.

\begin{figure}
\centering
\includegraphics[width=0.54\textwidth,height=8.5cm]{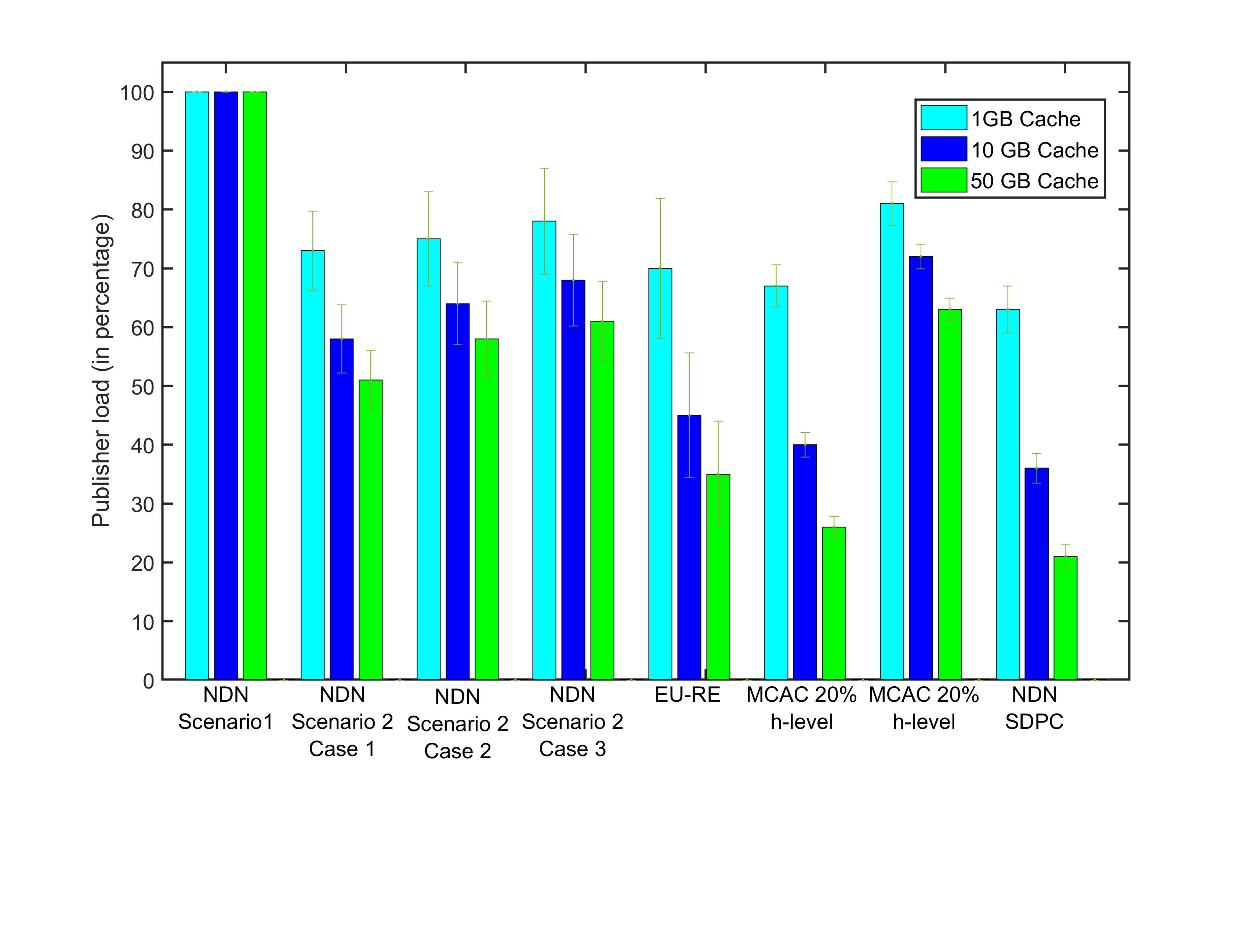}
\caption{ Publisher load for different cache sizes.}
\label{fig:7} 
\end{figure}

Figure~\ref{fig:7} shows the comparison of publisher load. We considered the case-3 level dynamicity of consumer space for EU-RE, MCAC, and NDN-SDPC.  From Figure~\ref{fig:7}, it is evident that in NDN-SDPC the publisher load is 12 to 20\% lower than EU-RE; however, publisher load at MCAC is almost same as NDN-SDPC, but MCAC's load increases with increasing number of h-level contents. This also implies that NDN-SDPC has higher cache hit ratio. 
\begin{figure}
\centering
\includegraphics[width=0.54\textwidth,height=8.5cm]{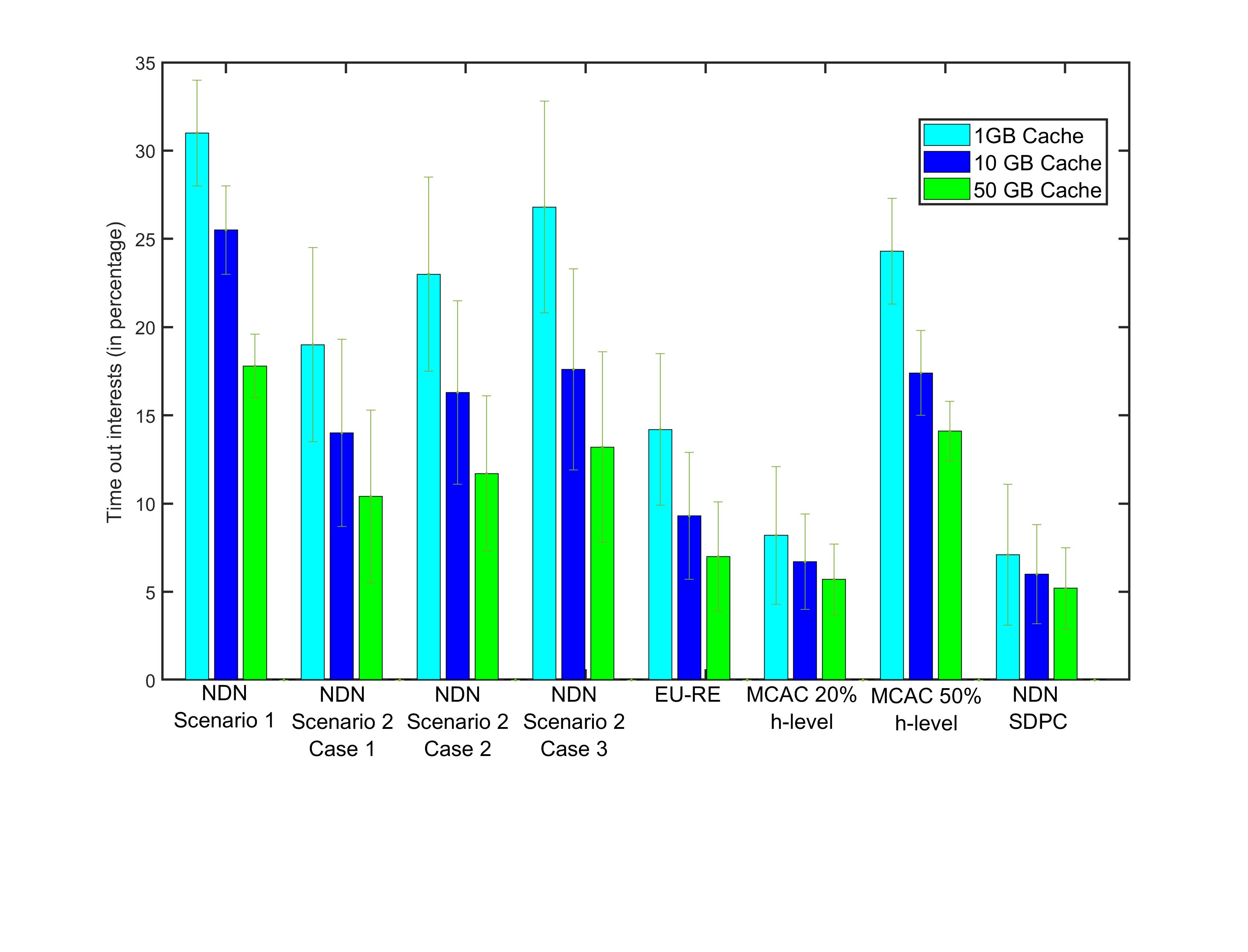}
\caption{Timeout interest ratio for different cache sizes.}
\label{fig:8} 
\end{figure}
Similarly, from Figure~\ref{fig:8}, it can be seen that EU-RE, NDN-SDPC and MCAC, with small numbers of h-level contents, suffer with lower number of time out interest packets. However, NDN-SDPC and MCAC with 20\% of h-level contents suffer 35 to 50\% less than EU-RE; however, this performance metric also shows that performance of MCAC reduces with increasing number of h-level contents. This also implies that in comparison to EU-RE and MCAC, the NDN-SDPC provides better cache diversity.

\section{Conclusion}
\label{sec:Conclusion}

For effective caching and access control of the protected content in ICN, we proposed a secure distribution of protected content (SDPC). The SDPC's keying protocol suite empowers the publisher and consumer to generate multiple symmetric encryption keys with the exchange of a single commitment key. Moreover, SDPC's subscription and content access protocol suite ensures that only authenticated users can acquire the respective key generation information for the requested content. Another important aspect of proposed scheme is the hybrid naming scheme, which provide privacy protection and deters the time analysis attack. The commitment key in SDPC is generated with the publisher’s public key, along with other secret credentials, and thus allows the consumer to implicitly verify the originality of the published content. In other words, self-certifying is achieved with the symmetric key cryptography, which makes SDPC free from the expensive computation overhead problem incurred in public key algorithms~\cite{koyama91}. Consequently, we believe that the adaptation of SDPC can make NDN more feasible for resource-limited networks such as Internet of things (IoT), which is one of our future works.

\appendices

\section{BAN logic Analysis}
\label{sec:appendixA}
Three basic objects of BAN logic are principals, formula/statements, and encryption keys. The principals and the protocol participants are represented by symbols \( P \) and \( Q \), respectively. The formula/statements are symbolized by \( X \) and \( Y \), and represents the content of the message exchanged. The encryption keys are symbolized by \( K \) . The logical notations of BAN-logic analysis are given in Table~\ref{tab:1}, and Some primary BAN-logic postulates used in the analysis of SDPC are summarized in Table~\ref{tab:2}.

\begin{table}
\centering
\caption{Logical notations of BAN-logic.}
\label{tab:1}       
\begin{tabular}{>{\centering}m{1.6cm}|m{6cm}}\hline{\smallskip}
Notation & Description   \\ \hline \hline

 \( P\models X\)   & \(P \) believes  \( X \) , or  \( P  \) would be enabled to believe  \( X \); in conclusion,  \( P \)  can take  \( X  \) as true.\\
 \hline
 \( P \mathrel{<\!\!\!\cdot} X \)   &  \( P  \) received a message  \( X  \) and  \( P \)  can see the contents of the message and is capable of repeating  \( X \)             \\
 \hline
 \( P \vert  \sim X \)  &  \(P \) has sent a message including the statement  \( X \). However, the freshness of message is unknown.            \\
 \hline
 \( P \Longrightarrow X\)  &  \(P \) controls \( X \)  and should be trusted for formula/statement  \( X \).         \\
 \hline
\( \# \left( X \right)  \)    &  \(X \) is\ fresh, i.e., \( X  \) never sent by any principal before.              \\
\hline
\( P \overset{k}{\leftrightarrow}Q  \)   &   \(P \) and \( Q \)  shares a key  \( K \) for secure communications and  \( K \) is only known to  \( P \) ,  \( Q \).             \\
\hline
\(  \left( X \right) _{K}  \)   &   The statement  \( X \)  is encrypted by key $K$.            \\
\hline
\(  \left( X \right) _{Y}  \)   &     It stand for  $X$ combined with $Y$. $Y$ is anticipated to be secret and its implicit or explicit presence proves the identity of a principal who completes $(X)_Y$.           \\
\hline
\noalign{\smallskip}           
\end{tabular}
\end{table}

\begin{table}
\centering
\caption{Primary BAN-logic postulates.}
\label{tab:2}       
\begin{tabular}{>{\centering}m{3.0cm}|m{4.5cm}}\hline{\smallskip}

Rule  &	Postulate   \\ \hline \hline
Message meaning rules  & $\frac{{P \models P\stackrel{K}{\longleftrightarrow}Q,P \mathrel{<\!\!\!\cdot} {{\left( X \right)}_K}}}{{P \models Q\mid  \sim X}},$ \newline
$\frac{{P \models P\stackrel{Y}{\longleftrightarrow}Q,P \mathrel{<\!\!\!\cdot} {{\left\{ X \right\}}_Y}}}{{P \models Q\mid  \sim X}}$ \\ 
 \hline
Nonce verification rule  & $\frac{{P \models \# \left( X \right),P \models Q\mid  \sim X}}{{P \models Q \models X}}$ \\
 \hline
Jurisdiction rule  & $\frac{{P \models Q \Rightarrow X,P \models Q \models X}}{{P \models X}}$\\
 \hline
Freshness rule  & $\frac{{P \models \# \left( X \right)}}{{P \models \left( {X,Y} \right)}}$\\
 \hline
Believe rule  & $\frac{{P \models Q \models \left( {X,Y} \right)}}{{P \models X,P \models Y}}$\\
 \hline
Session key rule  & $\frac{{P \models Q\# \left( X \right),P \models Q \models X}}{{P \models P \stackrel {K} {\longleftrightarrow} Q}}$\\
 \hline
\noalign{\smallskip}            
\end{tabular}
\end{table}

The SubP protocol should achieve the following four goals that state that both the consumer and the publisher trust the encryption key $K_S^i$ for the secure exchange of $KEY_{MSG}$:

\begin{itemize}
	\item G1: ${P_j} \models \left( {{N_i}\mathop  \leftrightarrow \limits^{K_S^i} {P_j}} \right)$
	\item G2:${N_i} \models \left( {{N_i}\mathop  \leftrightarrow \limits^{K_S^i} {P_j}} \right)$
	\item G3: ${P_j} \models {N_i} \models \left( {{N_i}\mathop  \leftrightarrow \limits^{K_S^i} {P_j}} \right)$
	\item G4:${N_i} \models {P_j} \models \left( {{N_i}\mathop  \leftrightarrow \limits^{K_S^i} {P_j}} \right)$
\end{itemize}

To verify the above-mentioned goals, the first step of BAN logic is to convert the subject protocol in its idealized form. The idealization is a process to represent each message exchange in its intended semantics. In other words, the idealization is a process of converting each message exchange into a logical formula by using BAN symbols and notations. The idealizations of SubP are given below.

\begin{itemize}

	\item M1: ${N_i}\xleftrightarrow{Via\,P_j}M:{\left\{ {\left( {{n_0},{N_i}\mathop  \leftrightarrow \limits^{n_S^i} M} \right)} \right\}_{K_{TS}^i = H\left( {K_p^j \oplus n_S^i} \right)}}$
	\item M2: ${P_j} \to M:\left\{ {{n_2},ID{P_j}} \right\}$
	\item M3:
\begin{align*}
& M \to {P_j}\colon \\
&\Biggl\{ 
\biggl[ n_1, n_2, \Bigl( 
  \textit{profile}, N_i,
  N_i(\xleftrightarrow{K_S^i})P_j,\\
  &\# \bigl( N_i(\xleftrightarrow{K_S^i}) P_j \bigr) 
  \Bigr)_{\!K_{\!P}^j} 
  \biggr]_{K_{\!P}^j}, \biggl[ n_0, n_1, \Bigl( 
  \textit{profile}, N_i,\\
&  N_i(\xleftrightarrow{K_S^i})P_j,
  \# \bigl( N_i(\xleftrightarrow{K_S^i})P_j \bigr)
  \Bigr)_{\!K_{\!P}^j},\quad N_i(\xleftrightarrow{K_S^i})P_j,\\
&  \# \bigl( N_i(\xleftrightarrow{K_S^i})P_j \bigr),
  N_i\xleftrightarrow{n_S^i}M 
  \biggr]_{K_{\mathit{TS}}^i = H(K_{\!P}^j \oplus n_S^i)}
\Biggr\}
\end{align*}

	\item M4: 
\begin{align*}
& {P_j}\xleftrightarrow{Via\,M}{N_i}\colon \\
&\Biggl\{  \biggl[ n_0, n_1, \Bigl(   \textit{profile}, N_i,\\
&  N_i(\xleftrightarrow{K_S^i})P_j,  \# \bigl( N_i(\xleftrightarrow{K_S^i})P_j \bigr)
  \Bigr)_{\!K_{\!P}^j},\quad N_i(\xleftrightarrow{K_S^i})P_j,\\
&  \# \bigl( N_i(\xleftrightarrow{K_S^i})P_j \bigr),  N_i\xleftrightarrow{n_S^i}M 
  \biggr]_{K_{\mathit{TS}}^i = H(K_{\!P}^j \oplus n_S^i)}, \\
&\biggl[  \left\langle {\zeta _0^j,{K_p}} \right\rangle ,\quad N_i(\xleftrightarrow{K_S^i})P_j,\\
&  \# \bigl( N_i(\xleftrightarrow{K_S^i})P_j \bigr),  N_i\xleftrightarrow{n_S^i}M  \biggr]_{K_{\mathit{TS}}^i = H(K_{\!P}^j \oplus n_S^i)}
\Biggr\}
\end{align*}

	\item M5:${P_j} \to M:{\left( {{n_1}} \right)_{K_S^i}}$
	\item M6:${N_i} \to {P_j}:{\left( {{n_1}} \right)_{K_S^i}}$ 
\end{itemize}

In an idealized protocol narration of SubP, the messages clearly show all the assertions. Using these assertions, all the implicit assumptions can be explicit. Then, the initial assumptions of SubP are given below.

\begin{itemize}
	\item A1:$M \models \# \left( {{n_0}} \right)$
	\item A2:$M \models \# \left( {{n_2}} \right)$
	\item A3: ${P_j} \models \# \left( {{n_1}} \right)$
	\item A4: ${N_i} \models \# \left( {{n_1}} \right)$
	\item A5: ${N_i} \models \left( {{N_i}\xleftrightarrow{K_{TS}^i = H\left( {K_p^j \oplus n_S^i} \right)}M} \right)$
	\item A6: $M \models \left( {{N_i}\xleftrightarrow{K_{TS}^i = H\left( {K_p^j \oplus n_S^i} \right)}M} \right)$
	\item A7:${P_j} \models \left( {{P_j}\mathop  \leftrightarrow \limits^{K_P^j} M} \right)$
	\item A8:$M \models \left( {{P_j}\mathop  \leftrightarrow \limits^{K_P^j} M} \right)$
	\item A9: $M \models {N_i} \models \left( {{N_i}\xleftrightarrow{K_{TS}^i = H\left( {K_p^j \oplus n_S^i} \right)}M} \right)$
	\item A10: ${N_i} \models M \models \left( {{N_i}\xleftrightarrow{K_{TS}^i = H\left( {K_p^j \oplus n_S^i} \right)}M} \right)$
	\item A11: $M \models {P_j} \models \left( {{P_j}\mathop  \leftrightarrow \limits^{K_P^j} M} \right)$
	\item A12: ${P_j} \models M \models \left( {{P_j}\mathop  \leftrightarrow \limits^{K_P^j} M} \right)$
\end{itemize}


Let us analyze SDPC to show that $N_i$ and $P_j$ share a session key. From M1, we have
\begin{equation}\label{Eq1}
M \mathrel{<\!\!\!\cdot} {\left\{ {{n_0},\left( {{N_i}\mathop  \leftrightarrow \limits^{n_S^i} M} \right)} \right\}_{K_{TS}^i = H\left( {K_p^j \oplus n_S^i} \right)}}.
\end{equation}

\noindent A6 and the message meaning rule infer that
\begin{equation}\label{Eq2}
M \models {N_i}\mid  \sim \left\{ {{n_0},\left( {{N_i}\mathop  \leftrightarrow \limits^{n_S^i} M} \right)} \right\}.	
\end{equation}

\noindent A1 and the freshness conjuncatenation comprehend that
\begin{equation}\label{Eq3}
M \models \# \left\{ {{n_0},\left( {{N_i}\mathop  \leftrightarrow \limits^{n_S^i} M} \right)} \right\}.
\end{equation}

\noindent Also, (\ref{Eq2}), (\ref{Eq3}), and the nonce verification rule deduce that
\begin{equation}\label{Eq4}
M \models \left\{ {{N_i} \models n_S^i,{n_0},\left( {{N_i}\mathop  \leftrightarrow \limits^{n_S^i} M} \right)} \right\}.
\end{equation}

\noindent Then, (\ref{Eq4}) and the believe rule infer that
\begin{equation}\label{Eq5}
M \models {N_i} \models \left( {{N_i}\mathop  \leftrightarrow \limits^{n_S^i} M} \right).
\end{equation}

\noindent From A2, (\ref{Eq5}), and the jurisdiction rule, it can be concluded
\begin{equation}\label{Eq6}
M \models \left( {{N_i}\mathop  \leftrightarrow \limits^{n_S^i} M} \right).
\end{equation}

\noindent This belief confirms that $M$ has received a message from a legitimate $N_i$. A2, A1, (\ref{Eq3}), and the freshness conjuncatenation comprehend that
\begin{equation}\label{Eq7}
M \models \# \left\{ {{n_0},{n_2},\left( {{N_i}\mathop  \leftrightarrow \limits^{n_S^i} M} \right)} \right\}.
\end{equation}

According to the nonce freshness, (\ref{Eq7}) proves that $M$ confirmed that $N_i$ is recently alive and running the protocol. Further, from (\ref{Eq4}) and (\ref{Eq7}), $M$ has guaranteed that the $N_i$ has been running the protocol, apparently with $M$. This also proves that $M$ and $N_i$ agree on the nonce values corresponding to all the nonce in M1 and M2. These three proven claims are known as aliveness, weak agreement, and non-injective agreement and defined in \cite{ Cremers06,Cremers12}.

From M3, we have
\begin{equation}\label{Eq8}
{P_j} \mathrel{<\!\!\!\cdot} {\left( {{n_1},{n_2},{{\left( {{N_i}\mathop  \leftrightarrow \limits^{K_S^i} {P_j},\# \left( {{N_i}\mathop  \leftrightarrow \limits^{K_S^i} {P_j}} \right),n_0^i} \right)}_{K_P^j}}} \right)_{K_P^j}}.
\end{equation}

\noindent A7, (\ref{Eq8}), and $n_2$ nonce verification rule deduce that
\begin{equation}\label{Eq9}
{P_j} \models M\mid  \sim \left\{ {{n_1},n_0^i,{N_i}\mathop  \leftrightarrow \limits^{K_S^i} {P_j},\# \left( {{N_i}\mathop  \leftrightarrow \limits^{K_S^i} {P_j}} \right)} \right\}.
\end{equation}

\noindent In addition, A3, (\ref{Eq9}), and the freshness conjuncatenation comprehend that
\begin{equation}\label{Eq10}
{P_j} \models \# \left\{ {{n_1},n_0^i,{N_i}\mathop  \leftrightarrow \limits^{K_S^i} {P_j},\# \left( {{N_i}\mathop  \leftrightarrow \limits^{K_S^i} {P_j}} \right)} \right\}.
\end{equation}

\noindent (\ref{Eq9}), (\ref{Eq10}), and the nonce verification rule infer that
\begin{equation}\label{Eq11}
{P_j} \models M \models \left\{ {{n_1},n_0^i,{N_i}\mathop  \leftrightarrow \limits^{K_S^i} {P_j},\# \left( {{N_i}\mathop  \leftrightarrow \limits^{K_S^i} {P_j}} \right)} \right\}.
\end{equation}

\noindent (\ref{Eq11}) and the believe rule comprehend that
\begin{equation}\label{Eq12}
{P_j} \models M \models \left( {{N_i}\mathop  \leftrightarrow \limits^{K_S^i} {P_j}} \right).
\end{equation}

The logic belief proves that $P_j$ is confident and believes that $K_S^i$ is issued by $M$; moreover, the freshness of the key from (\ref{Eq10}) also suggests that $M$ is alive and running the protocol with $P_j$ and $N_i$. Further, from (\ref{Eq9}), (\ref{Eq10}), and (\ref{Eq11}), $P_j$ has guaranteed that $M$ has been running the protocol, apparently with $P_j$. This also proves that $M$ and $P_j$ are also agree on the nonce values corresponding to all the nonce in M3. This concludes that $M$ and $P_j$ also satisfy the liveness, weak agreement, and non-injective agreement. Consequently, (\ref{Eq11}), (\ref{Eq12}), and the jurisdiction rule conclude G1, i.e., 
\begin{equation}\label{Eq13}
{P_j} \models \left( {{N_i}\mathop  \leftrightarrow \limits^{K_S^i} {P_j}} \right).
\end{equation}


\noindent  From M4, we have
\begin{equation}\label{Eq14}
{N_i} \mathrel{<\!\!\!\cdot} {\left\{ {{n_1},{N_i}\mathop  \leftrightarrow \limits^{K_S^i} {P_j},\# \left( {{N_i}\mathop  \leftrightarrow \limits^{K_S^i} {P_j}} \right),{N_i}\mathop  \leftrightarrow \limits^{n_S^i} M} \right\}_{K_{TS}^i = H\left( {K_p^j \oplus n_S^i} \right)}}.
\end{equation}

\noindent (\ref{Eq14}), A5, and the message meaning rule comprehend that
\begin{equation}\label{Eq15}
{N_i} \models M\mid  \sim \left\{ {{n_1},{N_i}\mathop  \leftrightarrow \limits^{K_S^i} {P_j},\# \left( {{N_i}\mathop  \leftrightarrow \limits^{K_S^i} {P_j}} \right)} \right\}.
\end{equation}

\noindent Then, ({Eq15}), A4, and the freshness conjuncatenation rule infer that
\begin{equation}\label{Eq16}
{N_i} \models \# \left\{ {{n_1},{N_i}\mathop  \leftrightarrow \limits^{K_S^i} {P_j},\# \left( {{N_i}\mathop  \leftrightarrow \limits^{K_S^i} {P_j}} \right)} \right\}.
\end{equation}

\noindent (\ref{Eq15}), (\ref{Eq16}), and the nonce verification rule deduce that
\begin{equation}\label{Eq17}
{N_i} \models M \models \left\{ {{n_1},,{N_i}\mathop  \leftrightarrow \limits^{K_S^i} {P_j},\# \left( {{N_i}\mathop  \leftrightarrow \limits^{K_S^i} {P_j}} \right)} \right\}.
\end{equation}

\noindent Then, (\ref{Eq17}) and the believe rule infer that
\begin{equation}\label{Eq18}
{N_i} \models M \models \left\{ {{N_i}\mathop  \leftrightarrow \limits^{K_S^i} {P_j}} \right\}.
\end{equation}

The logic belief proves that $N_i$ is confident and believes that $K_S^i$ is issued by $M$; moreover, the freshness of the key from (\ref{Eq16}) also suggests that $M$ is alive and running the protocol with $N_i$. Further, from (\ref{Eq15}), (\ref{Eq16}) and (\ref{Eq17}), $N_i$ has guaranteed that $M$ has been running the protocol, apparently with $N_i$ and $P_j$. This also proves that $M$ and $N_i$ are also agree on the nonce values corresponding to all the nonce in M4. This concludes that $M$ and $N_i$ also satisfies the liveness, weak agreement, and non-injective agreement. Conseuqently, (\ref{Eq17}), (\ref{Eq18}) and the jurisdiction rule conclude G2, i.e., 
\begin{equation}\label{Eq19}
{N_i} \models \left\{ {{N_i}\mathop  \leftrightarrow \limits^{K_S^i} {P_j}} \right\}.
\end{equation}


From M5, we have
\begin{equation}\label{Eq20}
{N_i} \mathrel{<\!\!\!\cdot} ID{S_j}.
\end{equation}

\noindent (\ref{Eq13}), (\ref{Eq19}), (\ref{Eq20}), and the meaning rule comprehend G3, i.e., 
\begin{equation}\label{Eq21}
{P_j} \models {N_i} \models \left\{ {{N_i}\mathop  \leftrightarrow \limits^{K_S^i} {P_j}} \right\}.
\end{equation}

From M6, we have 
\begin{equation}\label{Eq22}
{P_j} \mathrel{<\!\!\!\cdot} {n_1}.
\end{equation}

\noindent (\ref{Eq13}), (\ref{Eq19}), (\ref{Eq21}) and nonce verification rule deduce G4 of 
\begin{equation}\label{Eq23}
{N_i} \models {P_j} \models \left\{ {{N_i}\mathop  \leftrightarrow \limits^{K_S^i} {P_j}} \right\}.
\end{equation}

The logic belief proves that $N_i$ is confident and $P_j$ also believes that $K_S^i$ is issued by $M$. Moreover, (\ref{Eq5}), (\ref{Eq11}), and (\ref{Eq17}) prove that all communicating partners are confident that their communication partners exactly follow their roles in the protocol and exchange the intended messages in the intended order. This proven claim is known as non-injective synchronization and defined in~\cite{Cremers06}.

\begin{IEEEbiography}[{\includegraphics[width=1in,height=1.25in,clip,keepaspectratio]{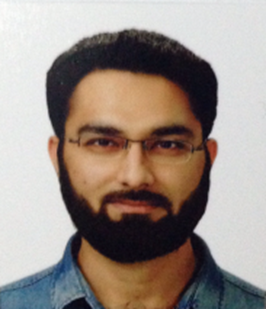}}]{Muhammad Bilal}(M'16) is an assistant professor of computer science in the Division of Computer and Electronic Systems Engineering at Hankuk University of Foreign Studies, Yongin, Rep. of Korea. He received his Ph.D. degree in Information and Communication Network Engineering from Korea University of Science and Technology, School of Electronics and Telecommunications Research Institute (ETRI) in 2017, M.S in computer engineering from Chosun University, Gwangju, Rep. of Korea in 2012, and B.Sc degree in computer systems engineering from University of Engineering and Technology, Peshawar, Pakistan in 2008. Prior to joining Hankuk University of Foreign Studies, he was a postdoctoral research fellow at Smart Quantum Communication Center, Korea University in 2017. He has served as a reviewer of various international Journals including IEEE Communications Magazine, IEEE Systems Journal, IEEE Access, IEEE Communications Letters, IEEE Transactions on Network and Service Management Journal of Network, IEEE IoT Journal, IEEE Transactions on Network Science and Engineering and Computer Applications, Personal and Ubiquitous Computing, and International Journal of Communication Systems. He has also served as a program committee member on many international conferences including IEEE VTC, IEEE IWCMC AI-IOT  and IEEE CCNC. He is an Editor of IEEE Future Directions Ethics and Policy in Technology Status and IEEE Future Directions Ethics and Policy in Technology Status. His primary research interests are Design and Analysis of Network Protocols, Network Architecture, Network Security, IoT, Named Data Networking, BlockChain, Cryptology and Future Internet.
\end{IEEEbiography}

\begin{IEEEbiography}[{\includegraphics[width=1in,height=1.25in,clip,keepaspectratio]{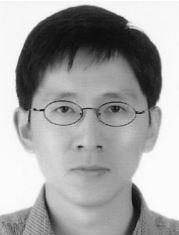}}]{Sangheon Pack} received the B.S. and
Ph.D. degrees from Seoul National University,
Seoul, South Korea, in 2000 and 2005, respectively. In 2007, he joined the faculty of Korea University, Seoul, Korea, where he is currently a Professor in the School of Electrical Engineering. He was the recipient of IEEE/Institute of Electronics and Information Engineers (IEIE) Joint Award for IT Young Engineers Award 2017, Korean Institute of Information Scientists and Engineers (KIISE) Young Information Scientist Award 2017, Korean Institute of Communications and Information Sciences (KICS) Haedong Young Scholar Award 2013, and IEEE ComSoc APB Outstanding Young Researcher Award. He served as a TPC vice-chair for information systems of IEEE WCNC 2020, a track chair of IEEE CCNC 2019, a TPC chair of EAI Qshine 2016, a publication co-chair of IEEE INFOCOM 2014 and ACM MobiHoc 2015, a co-chair of IEEE VTC 2010-Fall transportation track, and a publicity co-chair of IEEE SECON 2012. He is an editor of IEEE Internet of Things (IoT) Journal, Journal of Communications Networks (JCN), IET Communications, and he is a guest editor of IEEE Transactions on Emerging Topics in Computing (TETC). He is a senior member of the IEEE. His research interests include Future Internet, softwarized networking (SDN/NFV), information-centric networking (ICN)/delay tolerant networking (DTN), and vehicular networks. 
\end{IEEEbiography}

\end{document}